# Transfer-printed quantum-dot nanolasers on a silicon photonic circuit

Alto Osada,[1*] Yasutomo Ota,[1] Ryota Katsumi,[2] Katsuyuki Watanabe[1],

Satoshi Iwamoto[1,2], and Yasuhiko Arakawa[1,2]

[1]*Institute for Nano Quantum Information Electronics,*

*The University of Tokyo, Meguro 153-8505, Japan*

[2]*Institute of Industrial Science, The University of Tokyo, Meguro 153-8505, Japan*

*Corresponding author: alto@iis.u-tokyo.ac.jp

**Abstract**

Quantum-dot (QD) nanolasers integrated on a silicon photonic circuit are demonstrated for the first time. QD nanolasers based on one-dimensional photonic crystal nanocavities containing InAs/GaAs QDs are integrated on CMOS-processed silicon waveguides cladded by silicon dioxide. We employed transfer-printing, whereby the three-dimensional stack of photonic nanostructures is assembled in a simple pick-and-place manner. Lasing operation and waveguide-coupling of an assembled single nanolaser are confirmed through micro-photoluminescence spectroscopy. Furthermore, by repetitive transfer-printing, two QD nanolasers integrated onto a single silicon waveguide are demonstrated, opening a path to develop compact light sources potentially applicable for wavelength division multiplexing.

Silicon photonics [1,2] is a promising avenue towards the realization of highly-functional, low-cost photonic integrated circuits [3,4], owing to its compatibility with complementary-metal-oxide-semiconductor (CMOS) fabrication processes and the access to a variety of well-developed, densely-integrable micro/nanophotonic elements. However, the indirect bandgap of silicon hinders the straightforward implementation of efficient lasers on it. To overcome this issue, a tremendous amount of effort has been devoted to the hybrid integration of III-V lasers to the silicon platform, including those utilizing wafer bonding and direct crystal-growth techniques [5-22]. Whilst these approaches are known to be scalable, the minimum size of the hybridized III-V material is relatively large, which results in the waste of a large portion of III-V materials. Its size also hinders the facile and dense integration of diverse materials. In addition, the semiconductor micro/nanofabrication on the hybridized wafer tends to be complicated and usually require additional process optimization, which may delay the prototyping of the device chip. Resolving these issues become much more imperative for the integration of nanolasers, which utilize nanocavities for tightly confining light. Nanolasers are key elements in the pursuit of fast [23,24], high density [25,26] and energy-efficient operation [27,28] of light sources. Indeed, there has so far been a limited number of demonstrations on III-V nanolasers integrated on silicon [15,16], and thus providing the motivation to seek alternative integration approaches.

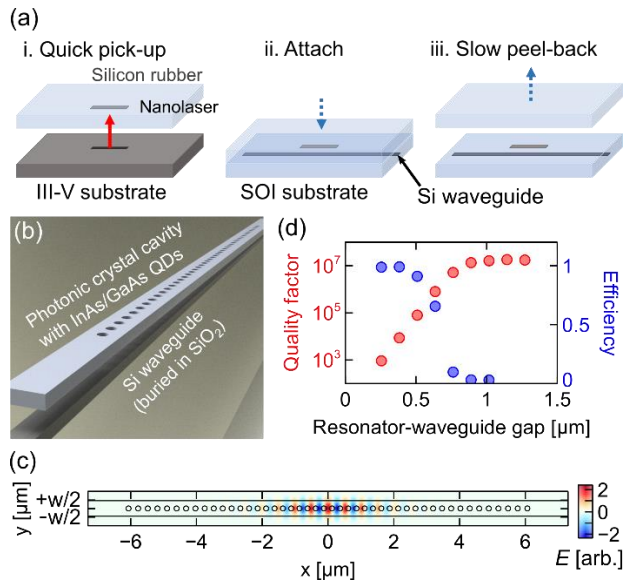

Fig. 1. (a) Schematic representation of the transfer-printing process. (b) Schematic illustration of a QD nanolaser coupled to a silicon waveguide. (c) Electric-field distribution of the fundamental mode of the photonic crystal nanocavity calculated by the FDTD method. The width of the nanobeam, $w$, is 472 nm. (d) Numerically evaluated quality factor (red) of the fundamental mode and the fraction of light coupled into a silicon waveguide (blue) as functions of the distance between the resonator and the waveguide.

In this context, the technique called transfer printing [29-31] is fascinating, since it allows for the hybrid integration of nanoscale elements by a simple pick-and-place assembly [see Fig. 1(a)], whereby various nanoscale structures can be independently fabricated, optimized and then integrated on silicon photonic circuits. Moreover, transfer printing enables area-selective material transfer to silicon substrate and hence efficient use of III-V material. It may also enable fast device prototyping in a way which is compatible with CMOS backend processing. So far, several types of transfer-printed nanolasers have been demonstrated [32-35], including the recent realization of a quantum-well nanolaser coupled to a silicon waveguide [35]. However, neither a single nor multiple nanolasers on a CMOS-processed silicon waveguide cladded by silicon dioxide has been realized using the transfer-printing method. In addition, there has been no report on a quantum dot (QD) nanolaser coupled to a silicon waveguide, despite the significance of QD for the realization of high-end nanolasers which are less prone to temperature variation [36], as well as feedback noise [37] that inherently exist in photonic integrated circuits.

In this Letter, we demonstrate InAs/GaAs QD nanolasers evanescently coupled to a CMOS-processed silicon waveguide by the transfer-printing method. The fabricated devices show clear features of lasing oscillation with high light coupling efficiencies (as high as 50%) into the silicon waveguide. We further employ transfer printing for assembling two nanolasers onto a single silicon waveguide. The two lasers exhibit multi-wavelength output from an output coupler, opening up a possibility to implement ultra-compact light sources useful for the wavelength division multiplexing. This work sheds light on an alternative pathway toward the realization of CMOS-compatible, densely integrated photonic integrated circuits illuminated with nanolasers.

First, we detail our design of the QD nanolaser coupled to the silicon waveguide buried in silicon dioxide, which is illustrated schematically in Fig. 1(b). The nanolaser is composed of a one-dimensional photonic crystal nanocavity (1D PhC) [38-42] made of GaAs containing InAs QDs. The thickness and the width of the 1D PhC nanobeam are 200 nm and 472 nm, respectively. The PhC contains 48 airholes of the diameter 152 nm, patterned with a lattice constant of 293 nm. From the PhC center toward the outside, we quadratically modulated the airhole period from 246 nm (center) to 293 nm (20th unit cell), resulting in the formation of defect modes in the 1D PhC. We analyzed the designed nanocavity by numerical simulations with the finite-difference-time-domain (FDTD) method. First, we characterized the 1D PhC nanocavity placed on silicon dioxide [38]. Figure 1(c) displays an electric-field distribution of the fundamental cavity mode, which resonates at the wavelength $\lambda$ of 1.17 μm with a quality factor of $1.8 \times 10^7$ and a mode volume of $0.82 \times (\lambda/n)^3$, where $n$ (=3.4) is the refractive index of GaAs. Next, we simulate the behavior of the 1D PhC nanocavity evanescently coupled to a glass-clad silicon waveguide. The width and thickness of the waveguide are 300 nm and 210 nm respectively, which allow for the single-mode operation around the wavelength of 1.17 µm. We controlled the coupling between the cavity and the waveguide by tuning the glass-filled gap between them, $d$. Figure 1(d) shows a summary of simulated quality factors and cavity-waveguide coupling efficiencies plotted as a function of $d$. As $d$ gets smaller, the quality factor of the cavity (red circles) degrades due to the enhanced coupling to the waveguide. This in turn improves the cavity-waveguide coupling efficiency (blue circles), which is defined by the ratio of light leakage from the cavity into the waveguide to the total light leakage [15]. For the gap of 370 nm, the cavity-waveguide system supports a near-unity coupling efficiency exceeding 99 %, together with a high quality factor of around $10^4$, which is sufficiently high for sustaining lasing oscillation with QD gain.

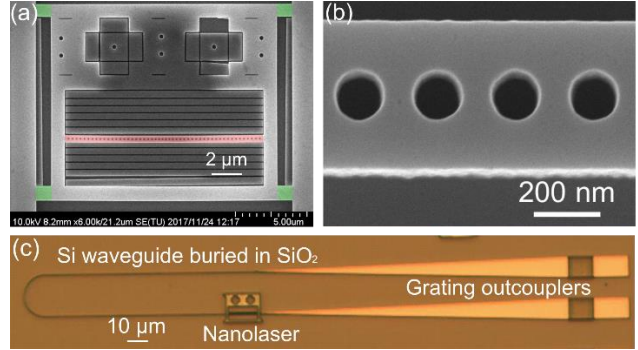

Fig. 2. (a) False-color scanning electron microscope image of a fabricated 1D PhC before being transferred onto a waveguide. A 1D PhC (red) is supported with a square frame suspended by four tethers (green). (b) Zoom-up image around the central part of the 1D PhC cavity. (c) Optical microscope image of a fabricated device, in which a 1D PhC is transferred onto a silicon waveguide.

Next, we describe the fabrication of the device. We fabricated 1D PhC nanocavities on a 200-nm-thick GaAs membrane embedding 6 layers of InAs QDs with an areal density of around $10^{10}$ /cm$^2$ per layer. The GaAs slab are grown on a 1-um-thick $Al_{0.3}Ga_{0.7}As$ sacrificial layer, which is later dissolved to form an airbridge structure suitable for transfer printing. We patterned the designed 1D PhC cavities through electron-beam lithography, dry and wet etching. Figure 2(a) displays a false-color scanning electron micrograph of a fabricated 1D PhC cavity before being transferred. The 1D PhC cavity (red) is suspended by a frame with four narrow tethers (green), which facilitates the sample pick-up during transfer printing. A close-up image of the nanobeam with airholes forming the 1D PhC is shown in Fig. 2(b). In parallel, we prepared silicon wire waveguides using a CMOS-process foundry. The silicon waveguides are buried in a silicon dioxide layer deposited by chemical vapor deposition with a thickness 2 µm. We thinned down the top cladding layer by dry etching and obtained a thickness of ~ 370 nm above the silicon waveguide. The root-mean-square surface roughness of the etched silica surface was measured to be 1.8 nm by an atomic force microscope, which was found to be sufficiently smooth for transfer printing.

The fabricated 1D PhC is then picked up and transfer-printed right above the buried, U-shaped silicon waveguide with the use of a transparent silicon rubber (Polydimethylsiloxane) [43], as schematically shown in Fig. 1(a). The sample pick-up and release are realized by controlling the adhesive force of the rubber stamp by tuning the peeling speed of the rubber [30]. The nanolaser is attached to the silicon photonic chip by van der Waals force. Nevertheless, the structure is mechanically rigid enough and supports a sufficient heat conductance for lasing [33]. During the transfer printing process, we monitored the positions of the nanocavity and the waveguide by an optical microscope. The relative position between them was controlled by a fine three-axis piezo stage and the alignment accuracy was found to be better than 100 nm by our transfer apparatus. An assembled device is shown in an optical micrograph in Fig. 2(c). The silicon waveguide is tapered toward two grating outcouplers, which vertically radiate light guided in the waveguide.

The fabricated device is evaluated by micro-photoluminescence (µPL) spectroscopy at 8 K. We perform optical carrier injection to the 1D PhC cavity ["Nanolaser" in Fig. 2(c)] by a pulsed titanium-sapphire laser of wavelength 808 nm, pulse width of 650 fs and repetition rate of 80 MHz. A 50x objective lens with a numerical aperture of 0.65 is used to focus the pump laser and to collect PL. The PL signals are analyzed by a spectrometer and a cooled InGaAs camera. We limit the collection area of PL signals by a spatial filter realized by narrowing the entrance slit of the spectrometer and the region of interest of the camera.

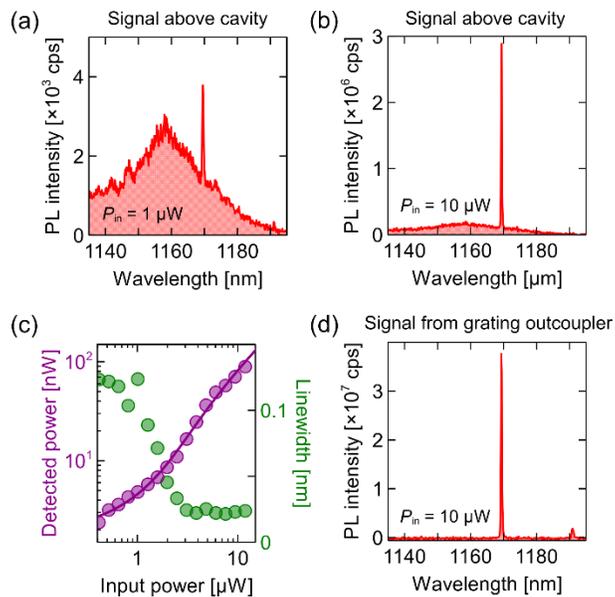

Fig. 3. (a,b) PL spectra observed above Nanolaser shown in Fig. 2(c) measured with input power being (a) $P_{in}$ = 1 µW and (b) $P_{in}$ = 10 µW. (c) Light-in-versus-light-out plot (purple) and measured linewidths (green) of the fundamental mode of the PhC nanocavity. Purple solid line shows a fitting result by a set of laser rate equations (see Appendix). (d) Spectrum obtained from one of the grating outcouplers recorded with an input power of $P_{in}$ = 10 µW.

First, we measure the PL from the nanocavity emitted into free space by collecting light just above the cavity. A spectrum measured with a weak irradiated pump power $P_{in}$ of 1 µW on average is shown in Fig. 3(a). The sharp peak at 1170 nm originates from the fundamental cavity mode, which becomes visible because of non-zero cavity-to-free-space radiation due to imperfect cavity-to-waveguide coupling. The quality factor of the cavity mode is evaluated to be $5 \times 10^3$, which seems to be limited by QD absorption and fabrication errors existing in the 1D PhC cavity. An intense broad peak at around 1160 nm arises from the spontaneous emission of the QDs. We also measure a PL spectrum at a higher pump power of 10 µW, as depicted in Fig. 3(b). The cavity signal has drastically increased its intensity and dominated the emission spectrum: the background spontaneous emission level is now lower than the cavity peak intensity by an order of magnitude. We investigated the pump power dependence of the cavity peak as summarized in Fig. 3(c). The measured light-in-versus-light-out (LL) plot (purple, left axis) exhibits a typical S shape, accompanied by a drastic increase of the output power with a threshold average pump power of 3 µW. In addition, we plot the evolution of cavity linewidth (green, right axis) which shows a significant narrowing around the lasing threshold. We observed the clamp of the QD spontaneous emission intensity at large pumping power as well (not shown). With these observations, the lasing operation of the fabricated device is confirmed. We further analyzed the measured LL curve by fitting using a laser rate equation model [44], as indicated by the purple solid line in Fig. 3(c). Through the fit, a spontaneous emission factor $\beta$, defined as the fraction of spontaneously emitted light that couples to the cavity, is extracted to be 0.2. The fitting also reveals the lasing threshold of 3.5 µW. The above analysis using the laser rate equations is detailed in Appendix.

When the PL radiated from the grating out-coupler is collected, the peak of the cavity mode is prevailing in the emission spectrum, as seen in Fig. 3(d). This observation suggests that only the cavity radiation efficiently couples to the waveguide, while the QD emission does not. At this point we can conclude that we succeeded in realizing the QD nanolaser coupled to the silicon waveguide. The coupling efficiency is estimated from the reduction of the quality factor of the cavity mode compared to those measured without the waveguide-coupling, $1.1 \times 10^4$, which yields the value of 55 %.

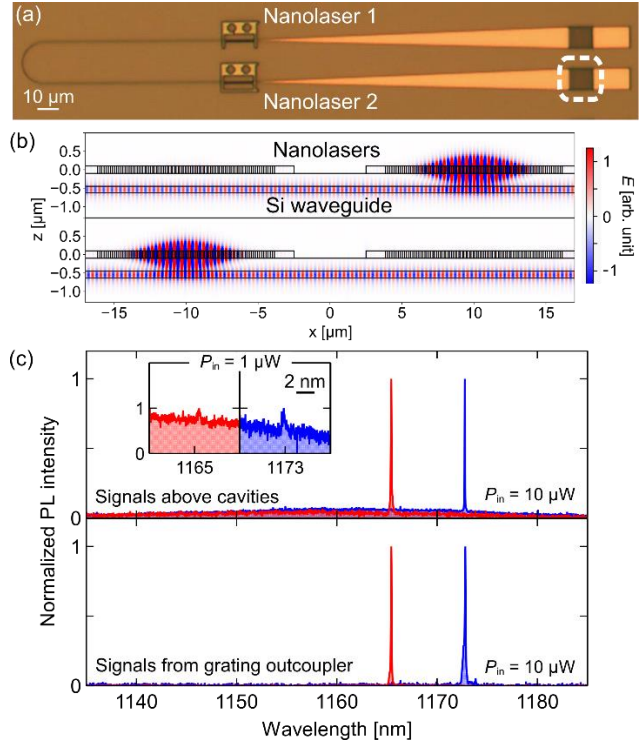

Fig. 4. (a) Optical microscope image of a pair of QD nanolasers printed on the same Si waveguide. (b) FDTD simulations demonstrating the independent light propagation of the cavity radiation in the waveguide. Note that the color scale is saturated within the cavity for the visibility of the waveguiding. (c) Emission spectra obtained above the nanolasers (top panel) and from the output coupler (bottom panel) under optical pumping to the Nanolaser 1 (red) and the Nanolaser 2 (blue).

As a further demonstration, we newly assembled two nanolasers with slightly different resonant wavelengths of ~ 10 nm on another single silicon waveguide by repeated transfer printing process. An optical microscope image of the fabricated device with the two nanolasers is displayed in Fig. 4(a). The two nanocavities do not interfere each other, confirmed numerically by the FDTD method, as shown in Fig. 4(b). The radiation from one of the two nanolasers propagates in the underlying silicon waveguide without re-entering or being scattered by the other nanolaser. µPL spectroscopy is performed by measuring the cavity-to-free-space leakage [top panel of Fig. 4(c)] or the radiation from one of the grating outcouplers [bottom panel]. For each detection scheme, we measure two spectra taken under respective optical carrier pumping into Nanolaser 1 (red) or 2 (blue) with an average power of $P_{in}$ = 10 µW. Whereas, the spectra in the inset of the top panel are recorded with $P_{in}$ = 1.0 µW. For all the measurement conditions, we observed a cavity emission peak originated from the fundamental cavity mode. The comparisons of the spectra in the top panel and the inset suggest abrupt increases of the cavity peak intensities with increasing pump power, together with the clamps of the QD spontaneous emission. Meanwhile, the spectra in the bottom panel imply the coupling of cavity radiation into the waveguide. We confirmed that the LL plots for the two nanolasers (not shown) also exhibit clear S-like shapes, the signatures of lasing operation. We fitted the LL curves by the laser rate equations and deduced the spontaneous emission factor of Nanolaser 1 (2) being $\beta = 0.080$ ($\beta = 0.11$). These observations verify the lasing oscillation of both nanolasers and their coupling to the silicon waveguide. The cavity-waveguide coupling efficiencies are estimated to be 30 % for Nanolaser 1 and 13 % for Nanolaser 2 by following the procedure described above. The reduced waveguide coupling efficiencies observed here presumably arise from the unwanted misalignment of the nanocavities with respect to the waveguide.

There are ways to improve experimental cavity-waveguide coupling efficiency, which is currently limited to 55 %. One is to increase the quality factor of the 1D PhC nanocavity on flat glass, which is now only ~ $10^4$ at the transparency condition and seems to be degraded mainly due to imperfections in the fabrication process of the 1D PhCs. Another way is to decrease the gap between the waveguide and the cavity to enhance the coupling between them. However, there is a tradeoff between the enhanced coupling and the increased lasing threshold power due to the degraded quality factor. In a situation that the quality factor of the 1D PhC cavity on flat glass is improved up to $10^5$, it is possible to achieve a coupling efficiency of 90% while keeping a sufficient quality factor of $10^4$ required for the lasing with the threshold around a few μW.

In conclusion, we demonstrated QD nanolasers evanescently coupled to a silicon waveguide. The heterogeneous integration of the nanolasers made of GaAs with InAs QDs onto the CMOS-processed silicon waveguide was realized by the transfer-printing technique. We obtained high cavity-waveguide coupling efficiencies as high as 55 %. Furthermore, two QD nanolasers are assembled on a single silicon waveguide by repeating the transfer process, leading to a compact multi-wavelength light source potentially applicable to the wavelength division multiplexing. The results presented in this paper manifest themselves as important stepping stones toward the future photonic integrated circuitry.

We acknowledge fruitful discussion with C. F. Fong. This work was supported by JSPS KAKENHI Grant-in-Aid for Specially Promoted Research (15H05700), KAKENHI 16K06294 and the New Energy and Industrial Technology Development Organization (NEDO).

**References**

1. D. Dai, J. Bauters and J. E. Bowers, Light: Science and Applications 1, e1 (2012).
2. R. Soref, IEEE J. Sel. Top. Quantum Electron. 12, 1678 (2006).
3. S. C. Nicholes, M. L. Mašanović, B. Jevremović, E. Lively, L. A. Coldren, and D. J. Blumenthal, Journal of Lightwave Technology 28, 641 (2010).
4. D. F. Welch et al., IEEE Journal of Selected Topics in Quantum Electronics 13, 22 (2007).
5. G. Roelkens, L. Liu, J. Brouckaert, J. Van Campenhout, F. Van Laere, D. Thourhout, and R. Baets, Mater. Today 10, 36 (2007).
6. A. W. Fang, H. Park, Y.-H. Kuo, R. Jones, O. Cohen, D. Liang, O. Raday, M. N. Sysak, M. J. Paniccia, and J. E. Bowers, Mater. Today 10, 28 (2007).
7. K. Tanabe, D. Guimard, D. Bordel, S. Iwamoto, and Y. Arakawa, Opt. Express 18, 10604 (2010).
8. K. Tanabe, K. Watanabe, and Y. Arakawa, Sci. Rep. 2, 1 (2012).
9. S. Chen, W. Li, J. Wu, Q. Jiang, M. Tang, S. Shutts, S. N. Elliott, A. Sobiesierski, A. J. Seeds, I. Ross, P. M. Smowton, and H. Liu, Nat. Photonics 10, 307 (2016).
10. Z. Wang, B. Tian, M. Pantouvaki, W. Guo, P. Absil, J. V. Campenhout, C. Merckling, and D. V. Thourhout, Nat. Photonics 9, 837 (2015).
11. A. Y. Liu, C. Zhang, J. Norman, A. Snyder, D. Lubyshev, J. M. Fastenau, A. W. K. Liu, A. C. Gossard, and J. E. Bowers, Appl. Phys. Lett. 104, 041104 (2014).
12. T. Wang, H. Liu, A. Lee, F. Pozzi, and A. Seeds, Opt. Express 19, 11381 (2011).
13. B. Jang, K. Tanabe, S. Kako, S. Iwamoto, T. Tsuchizawa, H. Nishi, N. Hatori, M. Noguchi, T. Nakamura, K. Takemasa, M. Sugawara, and Y. Arakawa, Appl. Phys. Express 9, 092102 (2016).
14. J. Kwoen, B. Jang, J. Lee, T. Kageyama, K. Watanabe, S. Iwamoto, Y. Arakawa, presented at the 44th International Symposium on Compound Semiconductor C2.4 (2017).
15. Y. Halioua, A. Bazin, P. Monnier, T. J. Karle, G. Roelkens, I. Sagnes, R. Raj, and F. Raineri, Opt. Express 19, 9221 (2011).
16. R. Chen, T.-T. D. Tran, K. W. S. Ko, L. C. Chuang, F. G. Sedgwick, and C. Chang-Hasain, Nat. Photonics 5, 170 (2011).

17. S. Keyvaninia, G. Roelkens, D. V. Thourhout, C. Jany, M. Lamponi, A. L. Liepvre, F. Lelarge, D. Make, G.-H. Duan, D. Bordel, and J.-M. Fedeli, Opt. Express 21, 3784, (2013).
18. G. Roelkens, L. Liu, D. Liang, R. Jones, A. Fang, B. Koch, and J. Bowers, Laser & Photonics Reviews 4, 751 (2010).
19. D. Liang and J. E. Bowers, Nat. Photonics 4, 511 (2010).
20. I.-S. Chung and J. Mørk, Appl. Phys. Lett. 97, 151113 (2010).
21. S. Matsuo, T. Fujii, K. Hasebe, K. Takeda, T. Sato, and T. Kakitsuka, Opt. Express 22, 12139 (2014).
22. G. Crosnier, D. Sanchez, S. Bouchoule, P. Monnier, G. Beaudoin, I. Sagnes, R. Raj, and F. Raineri, Nat. Photonics 11, 297 (2017).
23. M. Takiguchi, A. Yokoo, K. Nozaki, M. D. Birowosuto, K. Tateno, G. Zhang, E. Kuramochi, A. Shinya, and M. Notomi, APL photonics 2, 046106 (2017).
24. K. Takeda, T. Sato, A. Shinya, K. Nozaki, W. Kobayashi, H. Taniyama, M. Notomi, K. Hasebe, T. Kakitsuka, and S. Matsuo, Nat. Photonics 7, 569 (2013).
25. M. P. Nezhad, A. Simic, O. Bondarenko, B. Slutsky, A. Mizrahi, L. Feng, V. Lomakin, and Y. Fainman, Nat. Photonics 4, 395 (2010).
26. J. H. Lee, M. Khajavikhan, A. Simic, Q. Gu, O. Bondarenko, B. Slutsky, M. P. Nezhad, and Y. Fainman, Opt. Express 19, 21524 (2011).
27. M. Takiguchi, H. Taniyama, H. Sumikura, M. D. Birowosuto, E. Kuramochi, A. Shinya, T. Sato, K. Takeda, S. Matsuo, and M. Notomi, Opt. Express 24, 3441 (2016).
28. Y. Ota, M. Kakuda, K. Watanabe, S. Iwamoto, and Y. Arakawa, Opt. Express 25, 19981 (2017).
29. E. Menard, K. J. Lee, D.-Y. Khang, R. G. Nuzzo, and J. A. Rogers, Appl. Phys. Lett. 84, 5398 (2004).
30. M. A. Meitl, Z.-T. Zhu, V. Kumar, K. J. Lee, X. Feng, Y. Y. Huang, I. Adesida, R. G. Nuzzo, and J. A. Rogers, Nat. Materials 5, 23 (2006).
31. J. Yoon, S.-M. Lee, D. Kang, M. A. Meitl, C. A. Bower, and J. A. Rogers, Adv. Optical Mater. 3, 1313 (2015).
32. B. Guilhabert, A. Hurtado, D. Jevtics, Q.Gao, H. H. Tan, C. Jagadish, and M. D. Dawson, ACS Nano 10, 3951 (2016).

33. H. Yang, D. Zhao, S. Chuwongin, J-H. Seo, W. Yang, Y. Shuai, J. Berggren, M. Hammar, Z. Ma, and W. Zho, Nat. Photonics 6, 615 (2012).
34. A. Tamada, Y. Ota, K. Kuruma, J. Ho, K. Watanabe, S. Iwamoto, and Y. Arakawa, Jpn. J. of Appl. Phys. 56, 102001 (2017).
35. J. Lee, I. Karnadi, J. T. Kim, Y.-H. Lee, and M.-K. Kim, ACS Photonics 4, 2117 (2017).
36. Y. Arakawa and H. Sakaki, Appl. Phys. Lett. 40, 939 (1982).
37. Y. Arakawa and A. Yariv, IEEE J. Quantum Electron. 22, 1887 (1986).
38. P. B. Deotare, M. W. McCutcheon, I. W. Frank, M. Khan, and M. Lončar, Appl. Phys. Lett. 94, 121106 (2009).
39. E. Kuramochi, H. Taniyama, T. Tanabe, K. Kawasaki, Y.-G. Roh, and M. Notomi, Opt. Express 18, 15859 (2010).
40. Y. Zhang, M. Khan, Y. Huang, J. Ryou, P. Deotare, R. Dupuis, and M. Lončar Appl. Phys. Lett. 97, 051104 (2010).
41. Y. Gong, B. Ellis, G. Shambat, T. Sarmiento, J. S. Harris, and J. Vučković, Opt. Express 18, 8781 (2010).
42. R. Ohta, Y. Ota, M. Nomura, N. Kumagai, S. Ishida, S. Iwamoto, and Y. Arakawa, Appl. Phys. Lett. 98, 173104 (2011).
43. R. Katsumi, Y. Ota, M. Kakuda, S. Iwamoto, and Y. Arakawa, arXiv:1801.07915 (2018).
44. G. Bjork and Y. Yamamoto, IEEE J. Quantum Electron. 27, 2386 (1991).

**Appendix: Rate equation analysis**

We analyze the characteristics of the investigated nanolaser by using the following rate equations [44]:

$$\frac{dN_c}{dt} = \sigma p(t) - \gamma_c N_c - \beta \gamma_c N_p (N_c - N_0),$$

$$\frac{dN_p}{dt} = -\gamma_{ph} N_p + \beta \gamma_c N_c + \beta \gamma_c N_p (N_c - N_0).$$

Here $N_c$ and $N_p$ are numbers of excited carriers and photons in the cavity. The radiative decay rate of a carrier in a QD, $\gamma_c = 1.0 \text{ ns}^{-1}$, is used, whereas nonradiative decay processes are neglected here. The decay rate of the cavity mode is represented by $\gamma_p = 6.8 \text{ ps}^{-1}$ which is given by the quality factor of the cavity mode ~ $1.1 \times 10^4$. Transparency carrier number $N_0 = 150$ is estimated from the average number of QDs in the cavity region ~ 0.25 µm². The parameter $\beta$ reads the spontaneous emission factor and is treated as a fitting parameter. The term $\sigma p(t)$ describes the pumping of carriers by pulsed excitation with a pulse duration of 650 fs.